# User Centered and Ontology Based Information Retrieval System for Life Sciences

Sylvie Ranwez[1], Vincent Ranwez[2], Mohameth-François Sy[1],
Jacky Montmain[1], Michel Crampes[1]

[1] LGI2P Research Centre, EMA/Site EERIE, Parc scientifique G. Besse,
30 035 Nîmes cedex 1, France
firstname.lastname@mines-ales.fr
[2] Laboratoire de Paléontologie, Phylogénie et Paléobiologie,
Institut des Sciences de l'Evolution (UMR 5554 CNRS), Université Montpellier II,
CC 064, 34 095 MONTPELLIER Cedex 05, France
vincent.ranwez@univ-montp2.fr

**Abstract.** Because of the increasing number of electronic data, designing efficient tools to retrieve and exploit documents is a major challenge. Current search engines suffer from two main drawbacks: there is limited interaction with the list of retrieved documents and no explanation for their adequacy to the query. Users may thus be confused by the selection and have no idea how to adapt their query so that the results match their expectations.
This paper describes a request method and an environment based on aggregating models to assess the relevance of documents annotated by concepts of ontology. The selection of documents is then displayed in a semantic map to provide graphical indications that make explicit to what extent they match the user's query; this man/machine interface favors a more interactive exploration of data corpus.

## 1 Introduction

As the number of electronic data grows it is crucial to profit from powerful tools to index and retrieve documents efficiently. This is particularly true in life sciences where new technologies, such as DNA chips a decade ago and Next Generation Sequencing today, sustain the exponential growth of available data. Moreover, exploiting published documents and comparing them with related biological data is essential for scientific discovery. Information retrieval (IR), the key functionality of the emerging "semantic Web", is one of the main challenges for the coming years. Ontologies now appear to be a *de facto* standard of semantic IR systems. By defining key concepts of a domain, they introduce a common vocabulary that facilitates interaction between the user and the software. Meanwhile, by specifying relationships between concepts, they allow semantic inference and enrich the semantic expressiveness for both indexing and querying document corpus.

Though most IR systems rely on ontologies, they often use one of the two following extreme approaches: either they use most of the semantic expressiveness of the ontology and hence require complex query languages that are not really appropriate for non specialists; or they provide very simple query language that almost reduces the ontology to a dictionary of synonyms used in Boolean retrieval models [1]. Another drawback of most IR systems is the lack of expressiveness of their results. In most cases, results are simply proposed as a set of documents with no further explanations concerning the match between the documents and the query. Even when an IR system proposes a list of ranked documents, no explanation is provided with regard to (w.r.t.) this ranking, which means the result is not made explicit. In the absence of any justification concerning the results of IR, users may be confused and may not know how to modify their query satisfactorily in an iterative search process.

This paper describes an original alternative. Our system relies on a domain ontology and on entities that are indexed using its concepts (e.g. genes annotated by concepts of the Gene Ontology or PubMed articles annotated using the MeSH). It estimates the overall relevance of each entity w.r.t. a given query. The overall relevance of a document is obtained by aggregating the partial similarity measurements between each concept of the query and those indexing the document. Aggregation operators are preference models that capture end user expectations. The retrieved documents are ordered according to their overall scores, so that the most relevant documents (indexed with the exact query concepts) are ranked higher than the least relevant ones (indexed with hyperonyms or hyponyms of query concepts). More interestingly, defining an overall adequacy based on partial similarities enables a precise score to be assigned to each document w.r.t. every concept of the query. We summarize this detailed information in a small explanatory pictogram and use an interactive semantic map to display top ranked documents. Thanks to this approach, the end user can easily tune the aggregation process, identify, at a simple glance, the most relevant documents, recognize entity adequacy w.r.t. each query term, and identify the most discriminating ones.

The main objective of this work is to favor interactivity between end users and the information retrieval system (IRS). This interactivity is based on the explanation of how a document is ranked by the IR system itself: explaining how the relevance of a document is computed provides additional knowledge that is useful to end users to iterate their query more appropriately. This is achieved by evaluating how well each document matches the query based on both query/indexation semantic similarities and end user preferences and by providing a visual representation of retrieved entities and their relatedness relation to each query term.

In section 2 we review problems involved in information retrieval and describe the different approaches of similarity measurement used in this context. In section 3 we describe a new document-request matching model based on multi-level aggregations of relevance scores. In section 4 we illustrate the use of this approach for the identification of cancer genes and the interactive query rendering interface of our IRS. Finally, in section 5 we draw conclusions and look at future perspectives.

## 2    Information Retrieval: Overview of the State of the Art

The contribution of this paper is related to the use of semantics for information representation and visualization in information retrieval systems.

Information retrieval is generally considered as a sub-field of computer science that deals with the representation, storage, and access of information. The field has matured considerably in recent decades because of the increase in computer storage and calculus capacity and the growth of the World Wide Web. Some domains, such as life sciences, have particularly benefited from this technological advance. Nowadays, people no longer labor to gather general information, but rather to locate the exact pieces of information that meet their needs [2]. The main goal of an information retrieval system can thus be defined as "*finding material (usually documents) that satisfies an information need from within large collections (usually stored on computers)*" [3]. The main use of an IRS can thus be summarized as follows: needing information within an application context, a user submits a query in the hope of retrieving a set of relevant documents as the answer. To achieve this goal, IRSs usually implement three processes [2]:

- **The indexation process** aims to represent documents and queries with sets of (weighted) terms (or concepts) that best summarize their information content.
- **The search** is the core process of an IRS. It contains the system strategy for retrieving documents that match the query. An IRS selects and ranks relevant documents according to a score strategy that is highly dependent on their indexation.
- **The query expansion** is an intermediate process that reformulates the user query, based on internal system information, to improve the quality of the result.

In most IRSs, the indexation process boils down to representing both documents and queries as a bag of weighted terms (often called keywords) [4]. IRSs that use such document representation are keyword-based. A serious weakness of such systems is that they can be misled by the ambiguity of terms (e.g. homograph) and ignore relationships among terms (e.g. synonym or hyperonym) [5]. To overcome this difficulty, recent IRSs map keywords to the concepts they represent [6]. These concept-based IR systems thus need general or domain conceptual structures on which to map the terms. Conceptual structures include dictionaries, thesauri (Wordnet, UMLS) or ontologies (e.g. Gene Ontology). It is now widely acknowledged that their use significantly improves the performance of IRSs [7], and there is still room for improvement since most ontologies are not optimized to achieve this goal [8]. A survey of concept-based IR tools can be found in [6].

Many concept-based IRSs were developed based on theoretical frameworks for the indexing process as well as for relevance measurement [9]. The latter assigns a score to each document (called RSV – *Retrieval Status Value),* depending on how well it matches the query.

The work presented here is in line with the concept-based approach and takes as a starting point the existence of a domain ontology. Both documents and queries are represented by a set of concepts from this ontology. Fig. 1 gives an example based on the Gene Ontology to illustrate how ontologies can help reduce the number of relevant documents missed by IRSs (i.e. *silences*). Having the query: "*Organelle organization (GO_0006996)*" and "*Cardiac muscle fiber development (GO_0048739)*", the system may retrieve a document that has been indexed by concepts: "*Mitochondrion organization*" and "*Muscle fiber development*" as well as one (with smaller RSV) indexed by "*Cellular component organization*".

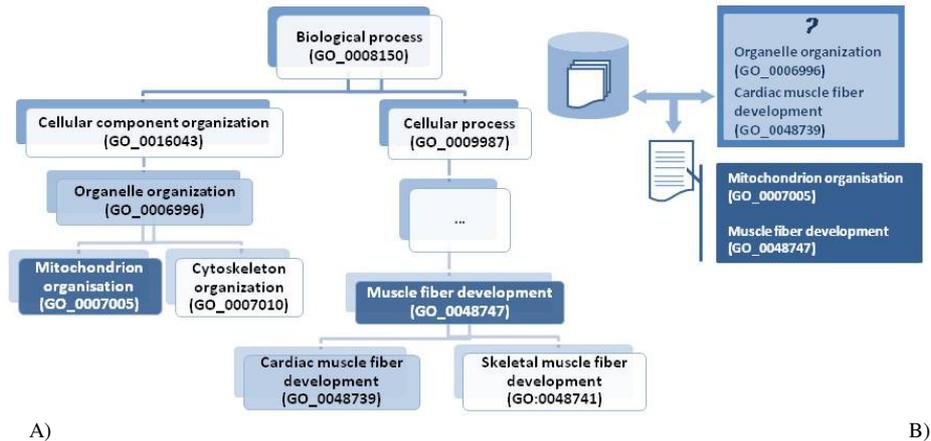

A)                                                                                                              B)

**Fig. 1** – How a domain ontology (A) can avoid *silences* in an IRS by expanding hyponyms and hyperonyms while querying an indexed corpus with a set of concepts (B).

### 2. 1   Boolean Request and their Generalizations

Boolean requests are certainly the most simple and widespread requests. However, studies indicate that even simple Boolean operators (AND, OR, NOT) are rarely used in web queries [10], and are even sometimes misused [11, 12]. Indeed, even when users know that all the terms must be included in the indexation (conjunctive request) or, on the contrary, that only one is needed (disjunctive requests), they do not mention it to the system. In the following, we thus focus on common requests where the user query is only a set of a few words.

Minkowski-Hölder's $L_p$ norms are aggregation operators that provide a theoretical framework to express whether a query is conjunctive or disjunctive using only one parameter [4]. They are particularly well suited to cases where the terms of the request are weighted. These weights may be related to term frequencies within the corpus, e.g. TF-IDF [4], or result from a fuzzy set indexation model. In this latter, a weight is associated with each concept indexing a document to represent to what extent a concept is a reliable indexation of a document [13].

Unfortunately, by summarizing the relevance of the document in a single score, aggregation operators tend to favor information loss and to fuzz out query results [14]. Indeed, unlike end users, they do not differentiate between documents whose scores result from cumulative minor contributions of all concepts within the query and those whose scores are due to the major contribution of a single concept. In addition, as they do not take advantage of semantic resources (ontologies, thesauri), they are unable to find relevant documents that are indexed by concepts that are different but semantically related to those of the query. Indeed, these operators only aggregate weights of a sub-set of terms: the ones that appear in the query. This statement is the basis of query expansion.

### 2. 2 Query Expansion

Query expansion is an intermediary step between the indexing and the matching process. As stated in [15], end users can rarely perfectly formulate their needs using query languages because they only have partial knowledge of IRS strategy, of the underlying semantic resources, and of the content of the database. Based on this statement, query refinement and expansion strategies have been developed to provide (semi-)automatic reformulation of the query. These reformulations may modify a query by adding concepts to it, by removing "poor" concepts from it or by refining the weights assigned to its query terms. Many query expansion (QE) techniques have been proposed, among which the widespread *relevance feedback* [16]. This query expansion technique uses the documents that are judged to be relevant by the user after an initial query to produce a new one using reformulation, re-weighting and expansion [17]. When done automatically, this process is called relevance back-propagation [18].

Query expansion may also be based on external vocabulary taken from ontologies or thesauri [19]. A common expansion strategy aims to supplement the query by adding its hyponyms. This method is an interesting complement to the Boolean search system detailed above. Indeed, it is then possible to select documents that are not indexed using exactly the same terms as the query and thus to avoid *silences*. This strategy is used for instance by the IRSs of PUBMED (http://www.ncbi.nlm.nih.gov /pubmed) and GOFISH [20]. However, since no distinction is made between the initial terms and those added users may be puzzled by the set of documents retrieved. Indeed, since they are not aware their query has been altered, they may not be able to understand the selection of a document indexed with none of their query terms.

### 2. 3 Semantic Similarity Measurements

It is possible to improve query expansion by using similarity measures. These measures not only enable selection of documents indexed with terms related to those of the query, but also retrieved documents to be ranked according to their semantic similarity to the query.

Since our approach extensively relies on semantic similarity measurements that significantly impact RSV calculus, we detail some of them below. As some of these measures satisfy distance axioms, we use semantic proximity, closeness or similarity randomly in the following.

The similarity measurements that have been proposed can be grouped in two main categories depending whether they are defined by *intention* or by *extension*. The first use the semantic network of concepts as metric space, and the second use a statistical analysis of term appearance in a corpus of documents [21].

While the semantic network may include various kinds of concept relationships, most intentional similarity measures only rely on the subsumption relationship, denoted as *is-a*, [22]. Indeed this relationship is the only one shared by all ontologies and it constitutes their backbone. The key role of the *is-a* relationship is clearly made explicit in the formal definition of an ontology proposed by [23] (p. 244-). The set of *is-a* relationships among concepts can be conveniently represented by an oriented graph whose vertices are concepts and whose edges indicate their subsumption relationship (*is-a*). Many concept similarities are based on this *is-a* graph. One of the most straightforward uses of this graph structure is to consider the length of the shortest path between two concepts $C_1$ and $C_2$ as their semantic distance [22]. If all the edges of the path have the same orientation, one concept is subsuming the other, but the more changes in direction the path contains, the harder it is to interpret. Therefore, [24] proposes to adapt this classical graph distance to produce a more sensitive proximity measurement, $\pi_{HO}(C_1, C_2)$, which takes into account the length of the path $P$ between $C_1$ and $C_2$, $lg(P)$ and the changes in direction within the path, $nbC(P)$:

$$\pi_{HO}(C_1, C_2) = \min_{P=(C1 \rightarrow C2)} \lg(P) + K * \text{nbC}(P) \qquad (1)$$

The K factor modulates the influence of changes in direction on the overall measurement. When K=0, $\pi_{HO}$ is equivalent to the distance proposed in [22]. On the other hand, a high value of K implies a minimum number of changes and thus a path that meets either the *least common ancestor* of $C_1$ and $C_2$, denoted by $lca(C_1,C_2)$ or one of their *greater common descendants*, denoted $gcd(C_1,C_2)$. Since 1994, when [25] first proposed to use *lca* in this context, it has played a key role in several similarity measurements. However, while focusing on the *lca*, this measurement neglects the symmetric notion of *gcd* and completely ignores whether concepts share common descendants, or not. [26] proposes a variant that takes this information into account.

One main limitation of all these graph-based measurements is that they assume edge homogeneity, whereas each edge of the *is-a* graph represents a specific degree of generalization or specialization. The semantic measurement proposed in [27] tries to capture this information based on the number of descendants of each concept. As this measurement is based on the *is-a* graph, it is denoted $d_{ISA}$ and authors demonstrated that it satisfies distance axioms. More formally, denoting by *hypo(C₁)* the set of hyponyms of $C_1$ (i.e. its descendants) and by *ancEx(C₁, C₂)* the set of concepts that are ancestors of either $C_1$ or $C_2$ (but not of both), $d_{ISA}$ is defined as:

$$d_{ISA}(C_1, C_2) = |hypo(ancEx(C_1, C_2)) \cup hypo(C_1) \cup hypo(C_2) - hypo(C_1) \cap hypo(C_2)| \qquad (2)$$

In this approach, the information content of a concept is evaluated by *intention* using only the ontology but not the corpus. Alternatively, *Extensional* measurements are mostly based on the corpus and often rely on the concept *information content* (or IC) defined in [21]. The IC of a concept $C_1$ is derived from the probability $P(C_1)$ that a document of the corpus is indexed by $C_1$ or one of its descendants:

$$IC(C_1) = -\log(P(C_1)) \qquad (3)$$

Combining the ideas of *lca* and *IC*, [21] introduces the notion of the most informative common ancestor (MICA) of a pair of concepts and defines a semantic proximity based on it as: $\pi_{Resnik} = IC(MICA(C_1, C_2))$. It should however be noted that MICA($C_1,C_2$) is not necessarily a *lca* of $C_1$ and $C_2$. This proximity measurement is tightly correlated with the individual *IC* of the two concepts. [28] proposes a variant to correct this bias:

$$\pi_{lin}(C_1, C_2) = \frac{2*IC(MICA(C_1,C_2))}{IC(C_1)+IC(C_2)} \qquad (4)$$

Proximities can be used in different contexts and their choice strongly depends on final objectives. Adequacy with *real* concepts' relatedness (i.e. the ones given by experts) must also be taken into account within the measurement choice [29, 30].

## 3 An Original Multi-level Score Aggregation to Assess Documents' Relevance Based on Semantic Proximity

Our work refers to concept-based IRSs. Our *Retrieval Status Values* (RSVs) are calculated from a similarity measurement between the concepts of an ontology. We propose to break down the RSV computation into a three stage aggregation process. First, we start with a simple and intuitive similarity measure between two concepts of the ontology (stage 1); then, a proximity measure is computed between each concept of the query and a document indexing (stage 2); finally, these measures are combined in the global RSV of the document through an aggregation model (stage 3). The last stage (aggregation) captures and synthesizes the user's preferences and ranks the collection of retrieved documents according to their RSV. The aggregation model enables restitution of the contribution of each query term to the overall relevance of a document. Hence it provides our system with explanatory functionalities that facilitate man-machine interaction and support end users in iterating their query. Furthermore in order to favor user interactions concept proximities must be intuitive (so that the end user can easily interpret them) and rapid to compute (so that the IRS is responsive even in the case of large ontologies).

We estimate the similarity of two concepts based on the Jaccard index between their descendant sets. Two main objectives are followed here: i) avoid *silence* when no document is indexed with the exact query concepts but with related concepts (hyponyms, hyperonyms) to increase the *recall* of the system; ii) make the query results more explicit concerning the way a match is computed, in particular documents indexed by query concepts and documents indexed by hyponyms or hyperonyms need to be distinguished.

### 3. 1  Semantic Similarity Between Concepts and Sets of Concepts

The choice of the semantic similarity measurement used by our IRS has a major impact on: i) the relevance of the retrieved documents, ii) the system's *recall* and iii) user comprehension of the document selection strategy. Hence, we chose a variant of the similarity measurement proposed by [28] (equation 3), with a valuation of the informational content of a concept based on the number of its hyponyms [31].

Because it has been emphasized that query concepts should only be replaced by hyponyms or hyperonyms, we estimate the semantic proximity of two concepts based on how much their hyponyms overlap (using the Jaccard index) as long as one is a hyponym of the other and otherwise we fix it at 0:

$$\pi_{JD}(C_1, C_2) = \begin{cases} \frac{hypo(C_1) \cap hypo(C_2)}{hypo(C_1) \cup hypo(C_2)} & si\ C_1 \in hypo(C_2)\ ou\ C_2 \in hypo(C_1) \\ 0 & on\ the\ other\ case \end{cases} \quad (5)$$

It should be noted that:
- $\pi_{JD}(C_1, C_1) = 1$
- $\pi_{JD}(C_1, C_2) < 1$, for each concept $C_1$ that is different from $C_2$
- $\pi_{JD}(C_1, C_2) > 0$, for each $C_1$ and $C_2$ having a hyponym relationship.

Several solutions have been proposed to extend similarity measurement between two concepts to measurement of similarity between two sets of concepts. This problem is of particular interest in life sciences because similarity between two gene indexations through the Gene Ontology (GO) may provide hints on how to predict gene functions or protein interactions [32]. Whereas comparing gene indexations (and document indexing in general) requires similarity measurements to be symmetric, this is not the case in IR. Indeed, when matching documents to queries, it seems normal to penalize a document because one concept of the query is absent from its indexing; on the other hand, penalizing a document because it is indexed by one concept absent from the query would be rather odd.

Given a similarity measurement between two concepts, the proximity between a query concept and a document can be defined as the maximum value of the similarities calculated between the query concept and each concept of the document indexing. This strategy leads to a simple and intuitive proximity measurement between each query concept and a document. More formally, if π denotes the similarity between two concepts from an ontology $O$, and $D_i$ denotes the $i^{th}$ concept of document $D$ indexing, $i = 1..|D|$, then we define the similarity between a concept $Q_t$ of the query and $D$ as $\pi(Q_t, D) = \max_{0 \leq i \leq |D|} \pi(Q_t, D_i)$.

### 3.2 Proximity Measurement Between a Document and a Query

After determining similarities between each concept of the query and (the index of) a document, the next step consists in combining them in a single score that reflects the global relevance of the document w.r.t. the query. User's preferences have to be taken into account during this process in order to determine the overall relevance of a document w.r.t. a query, i.e. its RSV.

As mentioned above, computing documents' RSV enables them to be ranked according to their relevance. Furthermore having the detail of the score of the document for each query concept allows us to justify and compare the source of the match of each document with the query. This is clearly related to the preference representation problem that has been extensively studied in decision theory [33]. A classical solution is to define a utility function $U$ in such a way that, for each alternative $D, D'$ in a list $\mathcal{D}$ of alternatives, $D \succeq D'$ (i.e. $D$ is preferred to $D'$) iff $U(D) \geq U(D')$. The decomposable model of Krantz [34] has been widely used when alternatives are $n$ dimensional. Following this model the utility function $U$ is defined as: $U(q_1,..,q_n) = h(u_1(q_1),..,u_n(q_n))$ with $u_t(.), t = 1..n$, are real-valued functions

User Centered and Ontology Based Information Retrieval System for Life-Science

in [0,1] and $h: [0,1]^n \to [0,1]$ is an aggregation operator that satisfies the following conditions:
- h is continuous;
- $h(0, 0,…, 0) = 0$ and $h(1, 1,…, 1) = 1$;
- $\forall (a_i, b_i) \in [0,1]^2$, if $a_i \geq b_i$ then $h(a_1, …, a_n) \geq h(b_1, …, b_n)$.

In our context, the *n* dimensional space corresponds to *n* query concepts. The *n* coordinates of a document correspond to its proximities with each concept of the query, i.e., $\pi(Q_t, D), t = 1..n$, defined in the previous section correspond to the $u_t(.)$ functions. The aggregation model combines the degrees of relevance (or matches) of a document indexing w.r.t. each query concept w.r.t. the user's preferences. The aggregation function *h* captures the preferences of the user: the way the elementary degrees of relevance are aggregated depends on the role of each query term w.r.t. the user's requirements. Three kind of aggregation can be distinguished:

- conjunctions (AND), $h(\pi(Q_1, D),..,\pi(Q_{|Q|}, D)) \leq \min_{t=1..|Q|} \pi(Q_t, D)$ ;

- disjunctions (OR), $h(\pi(Q_1, D),..,\pi(Q_{|Q|}, D)) \geq \max_{t=1..|Q|} \pi(Q_t, D)$ ;

- compromises $\min_{t=1..|Q|} \pi(Q_t, D) \leq h(\pi(Q_1, D),..,\pi(Q_{|Q|}, D)) \leq \max_{t=1..|Q|} \pi(Q_t, D)$ .

With the goal of improving man/machine interaction, we hope to give users a friendly and intuitive way of expressing their preferences concerning the overall relevance scoring strategy between a document and a query. We thus focus on compromise operators because they fit the widespread decision strategy that constrains the overall score to be between the minimum and the maximum value of elementary scores (convexity). Our approach is consequently based on Yager's operators [35]. These define a parameterized family of functions that represents compromise operators:

$$Y_m(\pi(Q_1, D),..,\pi(Q_{|Q|}, D)) = \left( \left( \sum_{t=1}^{|Q|} \pi(Q_t, D)^q \right) / |Q| \right)^{1/q}, q \in \mathbb{R} \qquad (6)$$

To get a better idea of the wide range of aggregation functions that are possible with this operators' family, let us exert some remarkable values:
- $q = 1$, arithmetic mean,
- $q = -1$, harmonic mean,
- $q \to 0$, geometrical mean,
- $q \to +\infty$, max (OR generalization)
- $q \to -\infty$, min (AND generalization)

A compromise operator can thus be selected by the user who may simply provide the value of parameter *q*. The choice of an aggregation operator is simply reduced to the choice of parameter *q* which still corresponds to our intuitive man/machine requirements. Indeed, our IRS interface includes a cursor to control the value of parameter *q* and to indicate whether the aggregation should tend toward a generalized "OR", a generalized "AND", or should tolerate more or less compensatory effects. When criteria do not play a symmetric role in the aggregation process, the relative importance of criteria can also be introduced in aggregation operators. In our case, it is possible to check that the Yager family can be extended to the weighted operators' family:

$$\overline{Y}_{wm}(\pi(Q_1,D),...,\pi(Q_{|Q|},D)) = \left(\sum_{t=1}^{|Q|} p_t .\pi(Q_t,D)^q\right)^{1/q} \quad (7)$$

However, in our application context, introducing weights can be more confusing than useful. Indeed, it is difficult for users to *a priori* assign weights to each of their query terms. Identifying precise values of weights requires specific procedures that are clearly thorny and cumbersome when simply writing a query. In this study, we thus focus on aggregation operators in which all the query terms play a symmetric role. Users only have to choose whether their compromise decisional behavior is closer to AND-like or OR-like.

This RSV 3-step computation (i.e. concept/concept, concept/document, query/document) has been integrated in an efficient and interactive querying system as detailed in the following section.

## 4   Results: OBIRS Prototype and Applications

Querying systems endowed with query expansion that add hyponym concepts to the query can be seen as the first step towards a semantic querying system. Our approach refines basic solutions to avoid *silences* by selecting documents that are indexed by the semantically closest hyponyms or hyperonyms of the query concepts. Furthermore, we are convinced that users should easily be able to understand the RSV at a glance to favor interaction with the IRS and query reformulation. Our 3-stage relevance model (which allows RSVs to be computed) integrates both the semantic expressiveness of the ontology based data structure and the end user's preferences. The more user friendly the man-machine interface, the more efficient the interaction between the IRS and the end-user.

To validate our approach, a corpus that contains the whole set of human genes referenced in the Ensembl database[1] has been chosen (~50.000 genes). Each gene is indexed with a subset of concepts of the Gene Ontology (about 30.000 concepts).

### 4. 1   Description of OBIRS and preliminary results

The screenshot presented in Fig. 2 shows our IRS (named OBIRS for Ontological Based Information Retrieval System) used here to find human genes indexed by Gene Ontology concepts. The query used for this screenshot is the same as the one used in section 2 (Fig. 1). The screen of the OBIRS interface is split vertically. On the left side, the querying interface provides assistance in expressing queries. Users are helped with the selection of query concepts (auto-completion, search concept with labels containing some terms) and can easily tune the aggregation function according to their preferences by moving a cursor from *rough* (strict conjunctive – "AND") to tolerant (disjunctive – "OR"). It is also possible to limit the number of documents retrieved (here 50) and to fix a threshold for the RSV (here 0.1).

---

[1] http://www.ensembl.org/

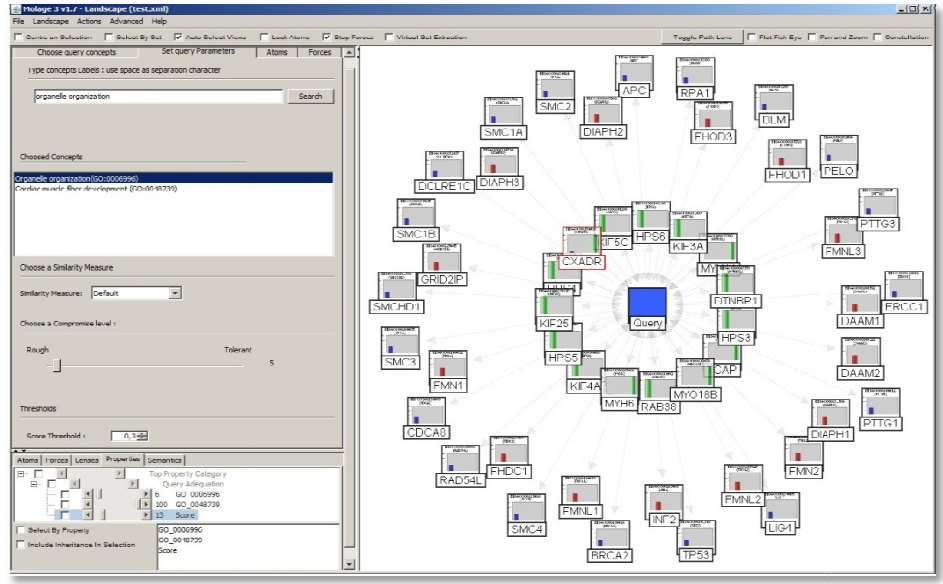

**Fig. 2 –** OBIRS interface: on the left, querying parameterization. On the right, display of selected documents – query = {*Organelle organization; Cardiac muscle fiber development*}.

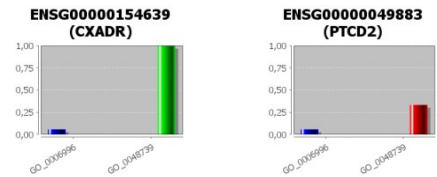

**Fig. 3 –** Examples of pictograms generated for genes CXADR and PTCD2 in response to the query {*Organelle organization; Cardiac muscle fiber development*}.

According to these parameters, the IRS selects relevant documents (human genes in this example) and for each document a histogram is generated to justify its match with the query. Each bar of a histogram is associated with a query concept and colored green, red or blue depending on whether the closest concept of the document indexing is exactly this query concept (green), a hyponym (red) or a hyperonym (blue). The size of the bar associated with a query concept $Q_t$ is proportional to the elementary relevance (i.e. $\pi(Q_t, D)$) of the document w.r.t. this concept (see Fig. 3). This information is detailed, when selecting a document, on the bottom left of the interface (here CXADR has been selected). These pictograms are then displayed on a semantic map in such a way that their physical distance to the query symbol (blue square in the middle of the screen) is proportional to the RSV of the document. Users can thus identify the most relevant documents at a glance and the reasons why they match the query, i.e., the contribution of each query concept to the RSV assessment. To facilitate the readability of the semantic map in Fig.2, a lens was designed that enlarges each pictogram in a popup in the proximity of the mouse (Fig.3).

Note that using the same query (*Organelle organization; Cardiac muscle fiber development*) the Ensembl database retrieves 0 genes with a classical Boolean strategy based on the "AND" operator and 15 genes based on a "OR". In OBIRS these 15 genes can easily be distinguished from additional ones (indexed with hyponym and hypernym concepts) since they are closer to the query symbol and their pictograms contain a green bar representing a perfect match (see Fig. 2). Hence, in OBIRS the *recall* is enhanced thanks to query expansion, and best genes can easily be identified through the visual display so that in practice there is no real loss of *precision*. We are working on further experiments to compare OBIRS *precision* and *recall* with other IRS and to estimate the influence of the semantic distance measurements on them.

### 4. 2   Application to Gene Identification

A request is built using the significantly over-represented GO terms of "molecular function" and "biological process" in cancer genes v.s. non-cancer genes (10 concepts of the table 1 in [36]). For a RSV threshold equals to 0.3 and a rather tolerant aggregation function (q=5.0), OBIRS proposes the genes that are shown in Fig. 4 (the higher the RSV the closer to the query symbol).

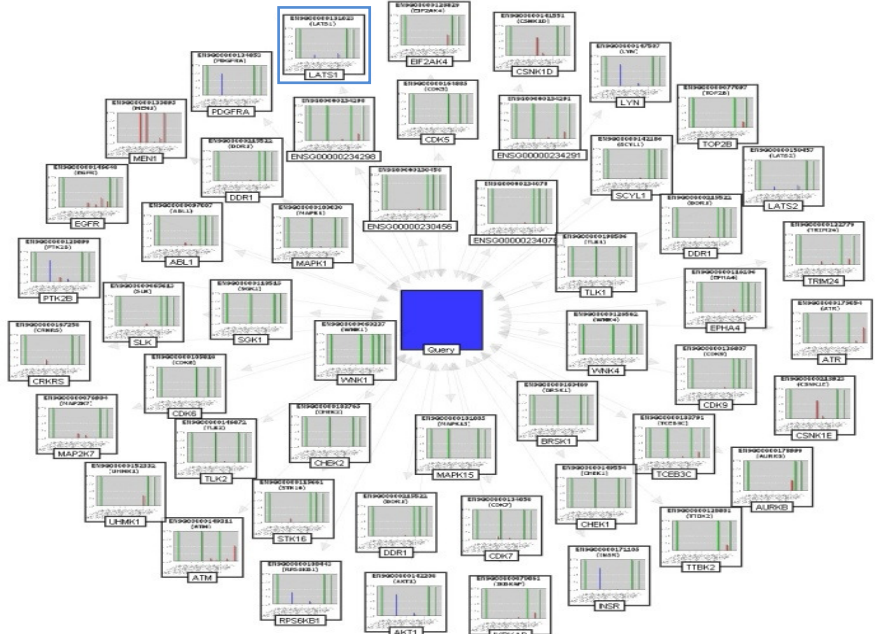

**Fig. 4** – Screenshot of OBIRS' 50 results w.r.t. the 10 over-represented GO-concepts (q= 5.0).

Several of these genes belong to the cancer genes listed in Cancer Genes Census[2] and most others are obviously also related to cancer (e.g. LATS1, framed in blue, stands for Large Tumor Suppressor). This query is processed in about one second on a standard desktop computer.

---

[2] http://www.sanger.ac.uk/genetics/CGP/Census/

When end-users consider that too many documents have been returned by the IRS, they can alter the relevance threshold: the lower the threshold, the stricter the selection. However, changing this threshold simply eliminates from the screen documents with a low RSV. But they can also modify the way the aggregation is performed (*rough/tolerant* cursor), and the semantic map is then completely reshaped because all the RSV are recomputed. As a result the closest documents in the second semantic map displayed may be completely different than the ones in Fig.4 (on Fig. 5, q=0.85).

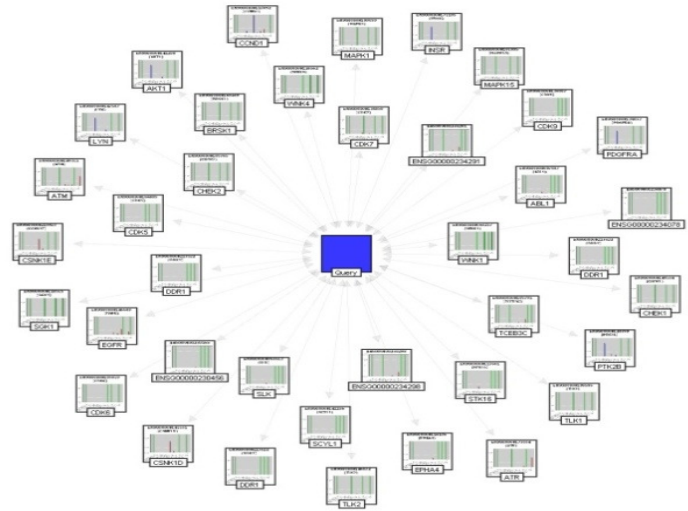

**Fig. 5 –** Same query as in Fig. 4 with q= 0.85 (less tolerant).

## 5    Conclusion and Perspectives

The approach described in this paper is an important step towards an IRS that benefits from the semantic expressiveness of ontologies while remaining easy to use. An original three stage aggregation model has been described to compute RSV scoring. This model has the particularity to embed end user preferences. The resulting OBIRS prototype is one of the first IRS able to elucidate its document selection to the user thanks to the decomposition of the RSV score that can be transcribed through intuitive pictograms. By locating these pictograms on a semantic map, OBIRS provides an informative overview of the result of the query and new possible interactions.

We are currently working on an OBIR extension that will let users reformulate their query through graphically selecting the documents they value and those in which they have no interest. This reformulation can be done by adding/removing concepts from the query, specifying/generalizing initial concepts of the query or adjusting the aggregation function. Reformulation leads to several optimization and mathematical questions but also raises important issues concerning feedback to users to enable them to continue to understand the IRS process and fruitfully interact with it.

We believe that there are many advantages to coupling the IR engine and rendering the result of the query, and that they should be considered simultaneously to provide a new efficient, interactive query environment. The RSV decomposition described in this paper is a good example of the benefit of simultaneously considering two related problems: i) how to rate documents w.r.t. a query ii) how to provide users feedback concerning rating of the documents. The latter is crucial to favor user/IRS intuitive interaction in iterative improvement of the query.

**Acknowledgments.** This work is the result of collaboration between ISEM (UMR 5554 – CNRS/UMII) and LGI2P-EMA. It was supported by the French Agence Nationale de la Recherche (ANR-08-EMER-011 "PhylAriane"). This publication is contribution No 2010-138 of ISEM.